\documentclass[aps, prd, letterpaper, 12pt, nofootinbib, superscriptaddress, longbibliography, notitlepage]{revtex4-1}
\usepackage[utf8]{inputenc}
\usepackage{amsmath,amssymb,amsfonts}
\usepackage{mathrsfs}
\usepackage{color}
\usepackage{graphicx} 

\definecolor{hgreen}{rgb}{0,.7,0}
\definecolor{hred}{rgb}{.7,0,0}
\definecolor{hblue}{rgb}{0,0,.7}

\usepackage[colorlinks=true,
linkcolor=hblue,
citecolor=hgreen,
filecolor=hblue,
urlcolor=hred]{hyperref}

\allowdisplaybreaks

\begin{document}

\title{Probing Lepton Flavor Violation at \texorpdfstring{\\}{} the ILC and CLIC \\ \small (Presented at the 32nd International Symposium on Lepton Photon Interactions at High Energies, Madison, Wisconsin, USA, August 25-29, 2025)}

\author{Pankaj~Munbodh}
\email{pmunbodh@ucsc.edu}
\affiliation{Department of Physics, University of California Santa Cruz, and
Santa Cruz Institute for Particle Physics, 1156 High St., Santa Cruz, CA 95064, USA}

\begin{abstract}
Lepton flavor violation in the $\tau\mu$ sector would be a clear sign of Beyond Standard Model physics. We employ the SMEFT framework to study the process $e^+e^-\to\tau\mu$ at the ILC and CLIC. We find that the $e^+e^-$ beam polarizations achievable at these machines allow us to probe the chirality structure of the SMEFT operators. In addition, the high center of mass energy leads to a substantial increase in sensitivity to the four-fermion operators that rivals, and in some cases, surpasses tau decay projections from Belle-II.
\end{abstract}

\maketitle

\section{Introduction} \label{sec:intro}
Lepton Flavor Violation (LFV) is a smoking gun signature for Beyond Standard Model (BSM) physics as it is highly suppressed in the Standard Model (SM)~\cite{Lee:1977qz, Lee:1977tib,Marciano:1977wx, Petcov:1976ff}. We focus on heavy new physics with a UV scale above the center of mass energies of the colliders $\Lambda\gtrsim\sqrt{s}$, which we parametrize by the Standard Model Effective Field Theory (SMEFT)~\cite{Grzadkowski:2010es}. We compute the cross section of the LFV process $e^+e^-\to\tau\mu$ for arbitrary $e^+e^-$ beam polarizations, and include the impact of Initial State Radiation (ISR) on the signal. We compare the sensitivity projections of the ILC and CLIC to those obtained from the Belle-II experiment that is searching for the related process $\tau\to\mu e^+e^-$.

\section{Observables}
Three classes of operators contribute to $e^+e^- \to \tau\mu$ at tree-level: dipole operators, Higgs current operators and four-fermion operators. The total cross section is given by

\begin{multline} \label{eq:sigma_tot}
\sigma(e^+e^- \to \tau\mu)
= \frac{m_Z^2}{64 \pi \Lambda^4} \bigg[ (1 + P_+)(1 + P_-) \Big( I^{e^+_R e^-_R}(s) + \bar I^{e^+_R e^-_R}(s) \Big) + (1 - P_+)(1 - P_-) \Big( I^{e^+_L e^-_L}(s) + \bar I^{e^+_L e^-_L}(s) \Big) \\ + \frac{2}{3} (1 + P_+)(1 - P_-) \Big( 2 I_0^{e^+_R e^-_L}(s) + 2 \bar I_0^{e^+_R e^-_L}(s) + I_2^{e^+_R e^-_L}(s) + \bar I_2^{e^+_R e^-_L}(s)\Big) \\ + \frac{2}{3}(1 - P_+)(1 + P_-)\Big( 2 I_0^{e^+_L e^-_R}(s) + 2 \bar I_0^{e^+_L e^-_R}(s) + I_2^{e^+_L e^-_R}(s) + \bar I_2^{e^+_L e^-_R}(s)\Big) \bigg] ~.
\end{multline}
where the $I^{e^+e^-}$ coefficients depend on the SMEFT coefficients and the $e^+e^-$ helicities. $P_-$ characterizes the polarization of the electron beam and $P_+$ characterizes the polarization of the positron beam. For detailed calculations as well as expressions for the coefficients, see Refs.~\cite{Altmannshofer:2025nbp, Altmannshofer:2023tsa}. An important qualitative feature is the linear growth of the cross section with $s$ in the presence of four-fermion operators.

\section{Expected sensitivities at ILC and CLIC}
\label{sec:sensitivity}
\begin{figure}[t]
\centering
\includegraphics[width=0.46\linewidth]{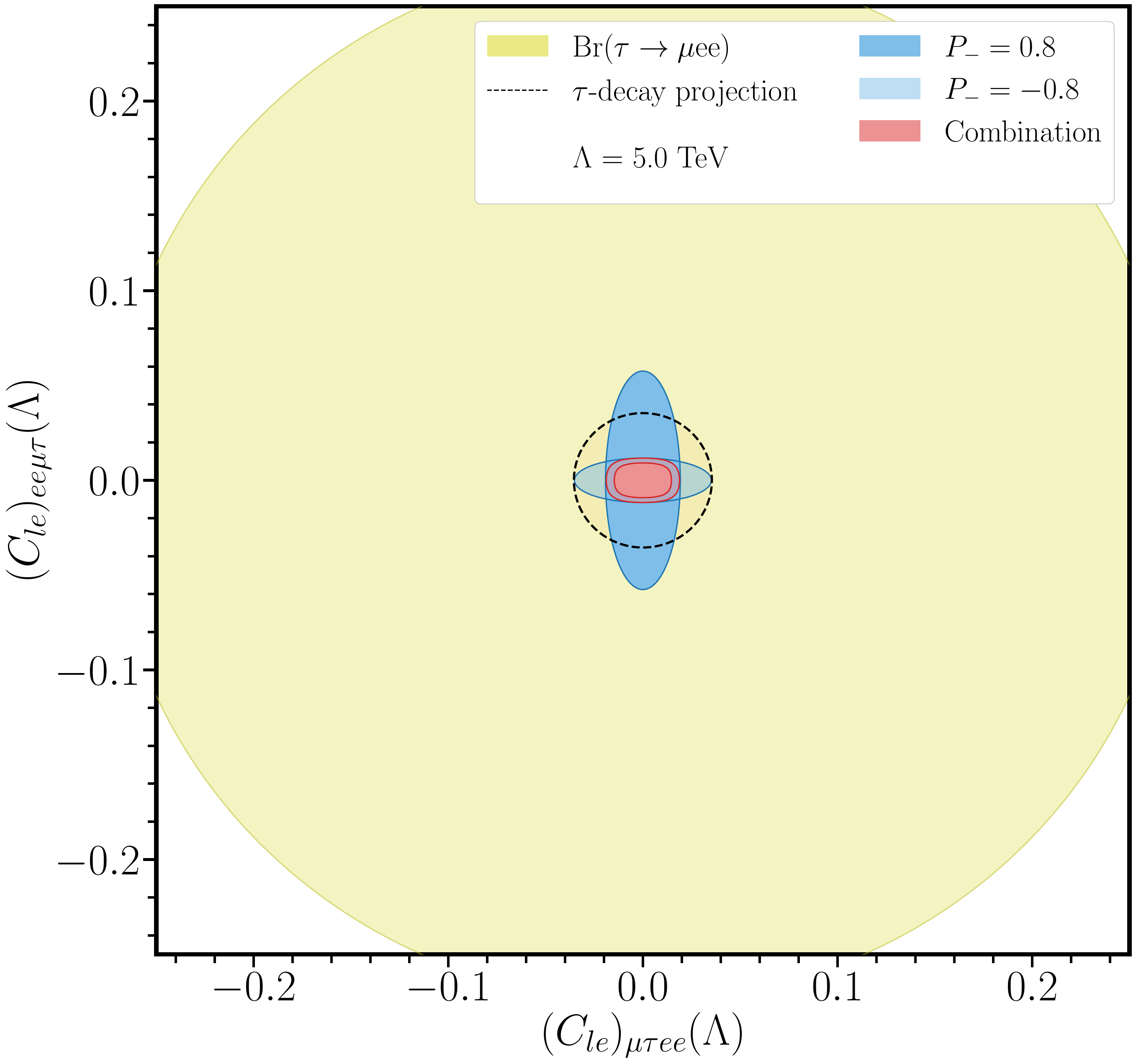} ~~~~
\includegraphics[width=0.46\linewidth]{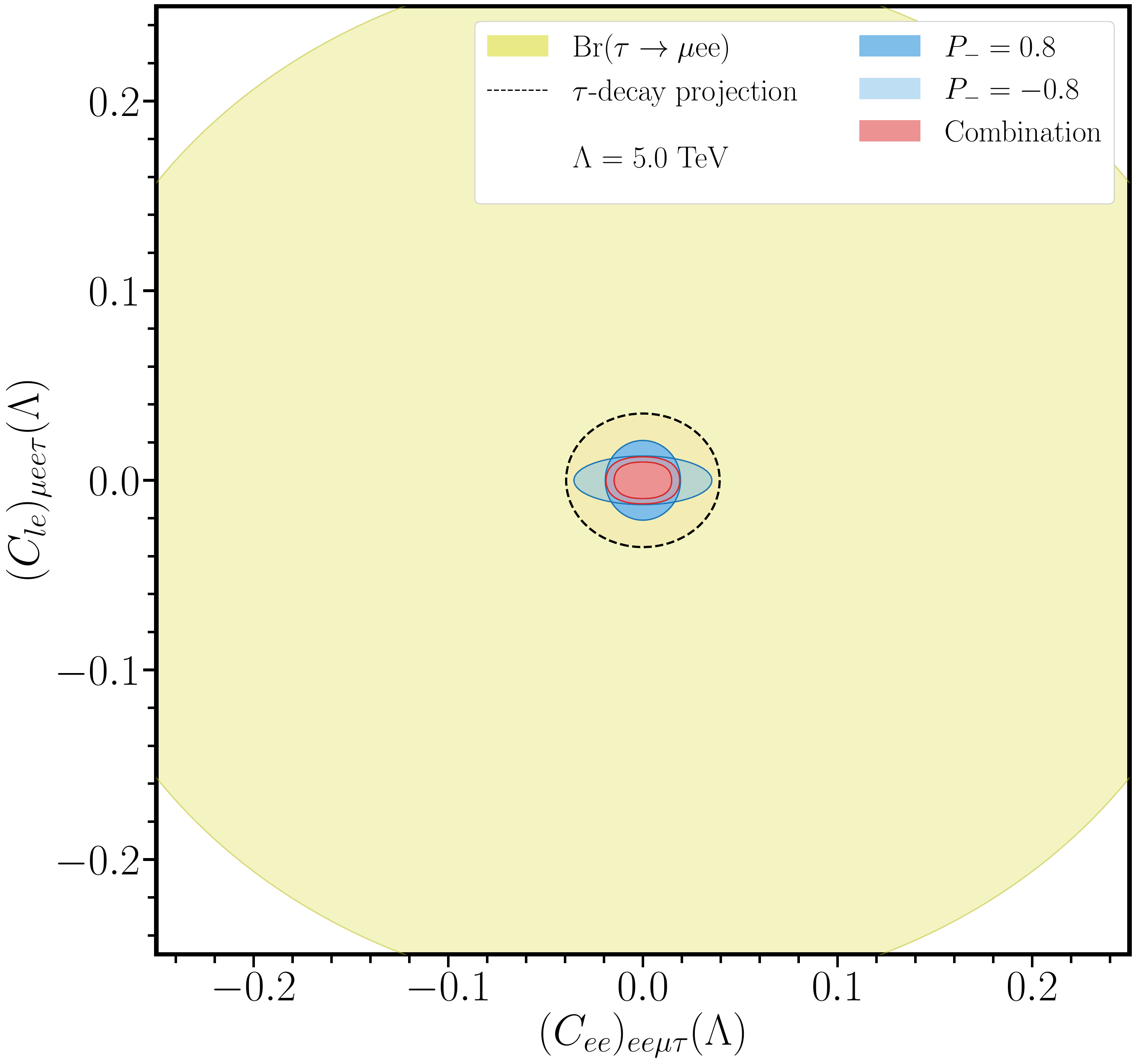}
\caption{Adapted from \cite{Altmannshofer:2025nbp}. Sensitivity to pairs of SMEFT coefficients at the $\sqrt{s} = 3$\,TeV run at CLIC. The various shades of blue represent our $2\sigma$ sensitivity projections for different $e^+/e^-$ beam polarizations in the process $e^+e^- \to \tau\mu$. The lighter and darker red regions correspond to the combined $2\sigma$ and $1\sigma$ constraints, respectively. The yellow region indicates the current $2\sigma$ bounds from searches for the $\tau\to\mu e^+e^-$ decay at BaBar and Belle~\cite{Hayasaka:2010np, BaBar:2010axs}, while the solid dashed line shows the expected sensitivity projection from Belle II~\cite{Belle-II:2018jsg}. The UV scale is fixed to $\Lambda = 5$ TeV.} 
\label{fig:Wilson_complementary_CLIC}
\end{figure}

\begin{figure}[!b]
\centering
\includegraphics[width = 0.92\linewidth]{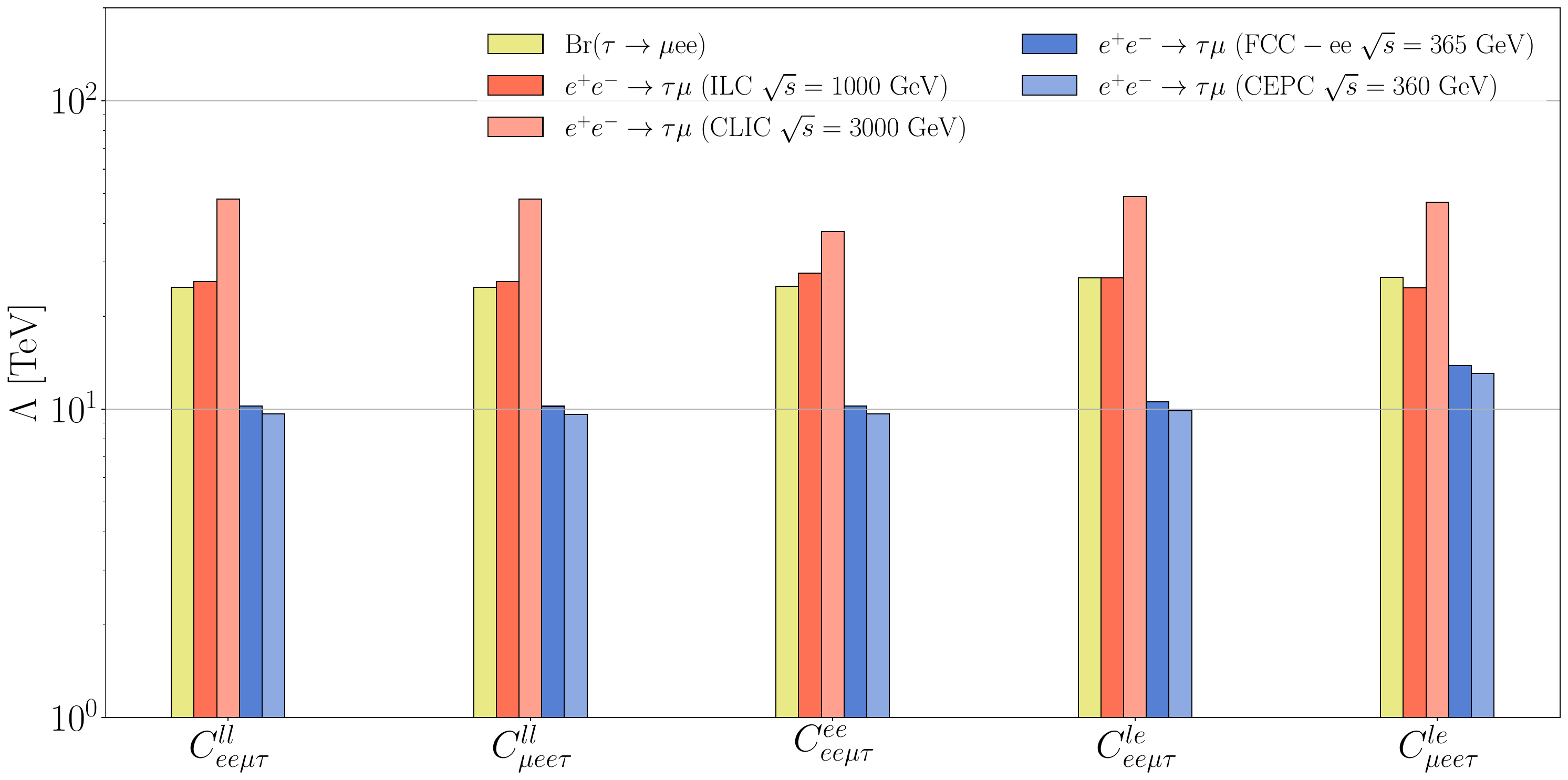}
\caption{Adapted from \cite{Altmannshofer:2025nbp}. Sensitivity to the new physics scale $\Lambda$ from $\tau\to\mu e^+ e^-$, and from the process $e^+e^-\to\tau\mu$ at future colliders. Each SMEFT Wilson coefficient is set to a value of unity at the scale $\Lambda$ ($C_i(\Lambda)=1$), with all others set to zero. Only projected sensitivities are shown.} 
\label{fig:bar_chart}
\end{figure}

\underline{Signal Simulation}
The cross section is convoluted simultaneously with the ISR radiator functions for both the $e^+e^-$ beams~\cite{NICROSINI1987551}, and with Gaussian distributions to account for the beam energy spread and the detector momentum resolution. We use a Monte Carlo (MC) procedure that unfolds in four stages. First, we sample two beam momenta from Gaussian distributions with a mean of $\sqrt{s}/2$
and with a standard deviation corresponding to the beam energy spread. Second, we sample from the ISR radiator functions for the electron/positron to lose a longitudinal momentum fraction. Third, we sample the momentum of the emitted muon from the differential cross-section as a function of momentum $p$ in the lab frame
\begin{equation}
\label{eq:dsigma_dp_signal}
    \frac{d\sigma}{dp} =   \frac{4}{\sqrt{s}} \frac{1}{(x_--x_+)} ~ \frac{d\sigma}{d\cos\theta} \quad \text{for} ~ \text{min}(x_-,x_+) \frac{\sqrt{s}}{2} < p < \text{max}(x_-,x_+) \frac{\sqrt{s}}{2}~.
\end{equation}
Here, the momentum of the $e^-$ (or $e^+$) is $x_-\sqrt{s}/2$ (or $x_+\sqrt{s}/2$) and $\theta$ is the angle between the electron and the outgoing muon in the center of mass frame. Finally, we smear the muon momentum with the detector momentum resolution according to a Gaussian distribution.\\
\noindent
\underline{Backgrounds}
To diminish background from $e^+e^- \to \mu^+\mu^-$ where a muon is misidentified as a tau, we select only the hadronic tau decay channels $\tau_{\rm had} \to \rho \nu \to 2\pi\nu$, and $\tau_{\rm had} \to 3\pi\nu$. The remaining backgrounds due to $e^+ e^- \to W^+ W^- \to \tau_\text{had} \nu \mu \nu$, $e^+ e^- \to W^+ W^- \to \tau_{\rm had} \nu \tau \nu \to \tau_{\rm had} \mu 4\nu$ and $e^+ e^- \to ZZ \to \tau_{\rm had} \tau \nu\nu \to \tau_{\rm had}\mu 4\nu$ are suppressed by imposing a cut ($x\gtrsim 1$) on $x = p_\mu/p_{\rm beam}$, where $p_\mu$ is the absolute value of the muon momentum in the lab frame and $p_{\rm beam}$ is the beam momentum~\cite{Dam:2018rfz}. \\
\noindent
Beam energy spread and finite momentum resolution of the detector shifts the endpoint of the muon momentum from the process $e^+ e^- \to \tau_{\rm had} \tau \to \tau_{\rm had} \mu \bar{\nu} \nu $ to be slightly above $x=1$ leading to a diminished background (as opposed to an $O(1)$ portion of the signal) to survive after the cut. The number of expected signal and background events are given by
\begin{align}
    N_{\rm sig} &= \mathcal{L}_{\rm int} \times\sigma(e^+e^- \to \tau \mu) \times \mathcal{B} (\tau \to \text{had}) \epsilon_{\rm had} \times  \epsilon_{\rm sig}^{x} \times \epsilon_{\rm sig}^{\rm ang} ~, \\
    N_{\rm bkg} &= \mathcal{L}_{\rm int} \times \sigma(e^+e^- \to \tau^+ \tau^-) \times 2 \times \mathcal{B} (\tau \to \text{had}) \epsilon_{\rm had} \times \mathcal{B}(\tau \to \mu \nu\nu) \times \epsilon_{\rm bkg}^{x} \times \epsilon_{\rm bkg}^{\rm ang} ~,
\end{align}
where $\mathcal{L}_{\rm int}$ is the integrated luminosity, $\epsilon^x_{\rm sig/bkg}$ are the cut efficiencies, $\epsilon_{\rm sig/bkg}^{\rm ang}\simeq98\%$ are the angular efficiencies and $\mathcal{B} (\tau \to \text{had}) \epsilon_{\rm had}$ is the branching ratio to two or three pions multiplied by their respective identification efficiencies. The center-of-mass energies, polarizations, luminosities, beam energy spreads and detector momentum resolutions can be found in Refs.~\cite{ILCInternationalDevelopmentTeam:2022izu, Behnke:2013lya} for the ILC and in Refs.~\cite{CLIC:2018fvx, Brunner:2022usy, CLICdp:2018cto} for CLIC.\\
\noindent
\underline{Results}
The criterion $N_{\rm sig} \geq 2 \sqrt{N_{\rm bkg} + N_{\rm sig}}$ provides us with an estimate of the sensitivity at $ 2 \sigma$.
The main results are shown in Figures (\ref{fig:Wilson_complementary_CLIC}) and (\ref{fig:bar_chart}) with a description in the Figure captions.

\section{Conclusion}
We gain exceptional sensitivity to four-fermion operators due to their linear scaling with $s$ and the high center of mass energy of linear colliders, especially CLIC. The different polarizations of the $e^+e^-$ offers an additional handle that allows us to probe the chirality structure of the operators.

\section{Acknowledgements}
Work presented in these proceedings is based on Refs.~\cite{Altmannshofer:2025nbp, Altmannshofer:2023tsa}. For a related proceedings at the tau lepton 2023 workshop, see Ref.~\cite{Munbodh:2024shg}. P.M. thanks Wolfgang Altmannshofer for helpful comments. P.M. thanks Jure Zupan and all the other organizers of the conference for the invitation to present this work.
The research of P.M. is supported in part by the U.S. Department of Energy grant number DE-SC0010107 and by the Chancellor's Dissertation Year Fellowship at UC Santa Cruz. 

\begin{appendix}

\end{appendix}

\bibliography{bibliography}

\end{document}